\documentclass[twocolumn,showpacs,preprintnumbers,amssymb]{revtex4}
\usepackage{graphicx}
\usepackage{dcolumn}
\usepackage{amsmath}
\usepackage{amssymb}

\begin{document}

\title{On the kinetic equation of Linear fractional stable motion  and applications to modeling the scaling of intermittent bursts.}


\author{N. W. Watkins$^{1,2}$, D. Credgington$^{1,3}$, R. Sanchez$^4$, S. J. Rosenberg$^1$, \& S. C. Chapman$^{5,2}$}
\affiliation{$^{1}$Physical Sciences Division, British Antarctic Survey (NERC),
 UK}
\affiliation{$^{2}$also at the Kavli Institute for Theoretical Physics, Santa Barbara, USA}
\affiliation{$^{3}$now at London Centre for Nanotechnology, University College London, London, UK}
\affiliation{$^{4}$Fusion Energy Division, Oak Ridge National Laboratory, Oak 
Ridge  TN 37830, USA}
\affiliation{$^{5}$Centre for Fusion, Space and Astrophysics,
University of
 Warwick, UK}
\date{\today}
 
\begin{abstract}
L\'{e}vy flights and fractional Brownian motion (fBm) have become   exemplars of  the heavy tailed jumps and long-ranged memory widely seen in physics. Natural time series frequently combine both effects, and  linear fractional stable motion (lfsm) is a model process of this type, combining $\alpha$-stable jumps with a memory kernel.
In contrast complex physical spatiotemporal diffusion processes where both the above effects compete have for many years been modelled using the fully fractional (FF) kinetic equation for the continuous time random walk (CTRW), with power laws in the pdfs of both jump size and waiting time.  We derive the analogous kinetic equation for lfsm and show that it has a diffusion coefficient with a power law in time rather than having a fractional time derivative like the CTRW.  We discuss some preliminary results on the scaling of burst ``sizes" and ``durations" in lfsm time series, with applications to modelling existing observations in space physics and elsewhere.
\end{abstract}
\preprint{NSF-KITP-08-81}
\maketitle

\section{Introduction}
 
Fractional kinetics is finding  increasingly wide application to physics, chemistry, biology and interdisciplinary complexity science\cite{KlagesBook,KlafterEA1996,Scalas2004,ZaslavskyBook,BalescuBook,Mura2008}. One reason for this is the link between ``strange" kinetics and observed non-Brownian anomalous diffusion, motivating the use of fractional dynamical models of transport processes, including those based on fractional calculus. \citet{Scalas2004} give numerous applications; we will just note  space plasmas \citep{MilovanovZelenyi2001}, magnetically confined laboratory plasmas \citep{Balescu1995,Carreras2001,CastilloNegrete2005,SanchezEA2005}, fluid turbulence \citep{KlafterEA1996} and the travels of dollar bills \cite{BrockmannEA2006}.  

Equally widespread in its application, and evolving in parallel with the theory of anomalous diffusion, is the theory of anomalous time series.  The corresponding models, particularly in the mathematics and statistics literature, have often been based on stable self-similar processes \cite{SamTaqBook,EmbrechtsMaejima2002,Sottinen2003}. Stability here means the property whereby  the shape of a probability density function (pdf) remains unchanged under convolution to within a rescaling (c.f. chapter 4 of \citet{MantegnaStanleyBook}) It is an  attractive feature in modelling, particularly when one anticipates that a signal represents a sum of random processes. In particular stable self-similar processes, a development in the wider field of stochastic processes\cite{Kolmogorov1956,Billingsley1979}, can model two effects which are are often seen in real data records. The first-Mandelbrot's ``Noah" effect \cite{Mandelbrot1963}- describes non-Gaussian ``heavy-tailed" pdfs, while the second-his ``Joseph" effect \cite{MandelbrotRSbook}-manifests itself as long-ranged temporal memory. The many applications have included hydrology \cite{MandelbrotVanNess1968}, finance \cite{Mandelbrot1963},  magnetospheric activity as measured by the auroral indices \cite{Watkins2002,WatkinsEA2005}, in-situ solar wind quantities \cite{WatkinsEA2005}, and solar flares \cite{BurneckiEA2008}. 

The existence of two rich, parallel, but intersecting literatures means that it is not yet completely known which techniques from one  will apply to a given problem in the other. It is for example not always clear {\it a priori} what type of kinetic equations will apply in a given context. The right class of kinetic equation for reversible microphysical  transport need not also be the right one  for an evolving time series taken from a macroscopic variable. The problem of model choice is an important and timely one, both in physics and more general complexity research. Because different models can predict subtly different observable scaling behaviours, distinguishing them may require measuring several exponents, as any individual exponent may be identical across several models, a point recently emphasised by \citet{Lutz2001}.

It is now increasingly recognised that much natural data is of the type that \citet{BrockmannEA2006} dubbed ``ambivalent". In such systems  heavy tailed jumps and long-ranged temporal memory compete to determine whether transport is effectively superdiffusive or subdiffusive. The ambivalent process they used for illustration, and  fitted to data, was the well-studied fully fractional continuous time random walk \cite{KlagesBook} which incorporates both  effects  via fractional orders of the spatial and temporal derivatives in its kinetic equation. \citet{ZaslavskyEA2007,ZaslavskyEA2008} also  advocated use of the same process in space physics  for modelling the auroral index time series. They  explicitly contested \cite{ZaslavskyEA2008} the applicability  of  a time series model \cite{WatkinsEA2005} based on a self-similar stable process-linear fractional stable motion (lfsm) \cite{SamTaqBook,EmbrechtsMaejima2002}-in this role. We  note that, rather than being purely a mathematical abstraction, lfsm has been linked to physics via the propagation of activity fronts in extremal models \cite{KrishnamurthyEA2000}. A comparison of these two approaches, leading to a better understanding of their structure and their similarities and differences, thus seems to us to be highly topical. It will be the first of two main topics of this Paper. Although we are fully aware that the kinetic equation we obtain on its own cannot fully specify a non-Markovian process, and, importantly, will not be unique to lfsm, we nonetheless believe that our comparison of the kinetic equations for the two paradigms is of value, particularly as  a source of physical insight (see also sections 1 and 2 of \cite{WatkinsEA2008}). 

To make the comparison we first briefly recap (Table 1) the main kinetic equations  corresponding to  the modelling of time series by stable processes, and of anomalous diffusion by the continuous time random walk (CTRW), respectively. In particular we highlight  (following \citep{Lutz2001})  the difference between fractional Brownian motion (fBm) and the fractional time process (ftp) which has sometimes led to confusion, at least in the physics and complexity literature (e.g. \cite{WatkinsEA2005}). We illustrate the potential value of this comparison with reference to a surprising gap in the physics literature, the absence of a kinetic equation corresponding to lfsm, analogous to the one given for fBm by \citet{WangLung1990}. We give a simple  derivation by direct differentiation using the characteristic function. The kinetic equation can be obtained by methods as diverse as a transformation $t^{\alpha H}$ of the time variable in the space fractional diffusion equation, and a path integral \cite{CalvoEA2008}. 

Our second main topic is  the potential relevance of lfsm to physics as a toy model for ``calibrating" diagnostics of intermittency  (c.f. \cite{MercikEA2003}). As a frequent attribute of nonequilibrium and nonlinear systems, intermittency has been a particular stimulus to physicists and time series modellers \cite{SornetteBook}. In particular the paradigm of self-organised criticality (SOC) \cite{SornetteBook} has been one framework for this, embodying the hypothesis of avalanches of activity in nonequilibrium complex systems. We investigate the scaling of intermittent bursts in lfsm, using the burst size and duration measures which have very often been used as direct diagnostics of SOC. Such measures have been previously studied on, among many others, magnetospheric  and solar wind time series \cite{FreemanEA2000}. We follow several earlier conjectures \cite{FreemanEA2000,Watkins2002, CarboneEA2004, Bartolozzi2007} and make simple scaling arguments building on a result of \citet{KearneyMajumdar2005} which suggest that lfsm could indeed be one candidate model for  observed power laws in such bursts.  We test our arguments with numerics using the algorithm of Stoev and Taqqu \cite{StoevTaqqu2004}. We confirm the earlier numerical results of  \citet{CarboneEA2004} and the more recent work of \citet{Rypdal2008}. These papers considered just the $\alpha=2$, fBm case, using a running average threshold and a fixed threshold respectively.  

However we find numerically that for lfsm our simple scaling argument, while giving a good approximation to the dependence of the burst size exponent on the self similarity exponent $H$ as $\alpha$ is progressively reduced from  2,  becomes much less accurate as $\alpha$ reaches 1. We conclude by offering some suggestions as to the reasons for this, and describing future work. 

\section{The kinetic equation of lfsm}

\subsection{Limit theorems and stochastic processes}

\begin{table}
\[
\begin{array}{*{20}c}
\hline
   &  \alpha  &\vline &  H  &\vline &  {{\rm{Stable \ process}}} & {{\rm{CTRW}}}   \\
& & \vline & & \vline & & \\
& & \vline & & \vline & H=1/\alpha+d& d^{'}= \alpha H\\
\hline
   &  {\alpha {\rm{ = 2}}} &\vline &  {H  = 1/2} &\vline &  {BWBm} &   &  \\
   &  {} &\vline &  {} &\vline &  \nabla ^2 p = \partial_t p & {}  &  \\
\hline
   &  {0 <  \alpha  \le 2} &\vline &  {H  = 1/\alpha} &\vline &  {oLm} &   &  \\
   &  {} &\vline &  {} &\vline &  \nabla ^\alpha  p = \partial_t p  & {}  &  \\
\hline
   &  {\alpha  = 2} &\vline &  0 \le H \le 1 &\vline &  {fBm} & {ftp}  &  \\
  &  {} &\vline &  {} &\vline &  2Ht^{2H-1} \nabla ^2 p = \partial_t p  & \nabla ^2 P = \partial_t^{\alpha H} p  &  \\
\hline
   &  {0 \le \alpha  < 2} &\vline &  0 \le H \le 1  &\vline &  {\bf lfsm} & {ap}  &  \\
   &  {} &\vline &  {} &\vline & {\bf  \alpha H  t^{\alpha H-1} \nabla ^\alpha  p = \partial_t p } & \nabla ^\alpha  p = \partial_t^{\alpha H} p  &  \\
\hline
\end{array}
\]
\caption{Kinetic equations for the main classes of process used to study anomalous time series and transport beyond the Bachelier Wiener Brownian paradigm. Ordinary L\'{e}vy motion (oLm) parameterised by a stability exponent $\alpha$ relaxes the finite variance assumption of the central limit theorem. The fractional time process (ftp), and the ambivalent process (ap) i.e. the fully fractional continuous time random walk, add a fractional derivative of order $\alpha H$ to the kinetic equations for WBm and oLm respectively.  A different way of introducing temporal memory effects is via stable selfsimilar processes with memory kernels, fractional Brownian motion (fBm)and linear fractional stable motion (lfsm) respectively. Although not fully specified by them, the stable processes nonetheless have kinetic equations with  time dependent diffusion coefficients. To our knowledge the (boldfaced) kinetic equation for lfsm had not been derived before our preprint \cite{WatkinsEA2008}. It was subsequently arrived at by path integral methods in \cite{CalvoEA2008}.  In the self-similar  stable processes the self similarity parameter $H$ depends on a memory exponent $d$ and on the stability exponent $\alpha$ via $H=1/\alpha+d$. In the CTRW case by contrast the standard memory exponent, $d'=\alpha H$. In all cases there is a coefficient $D$ with appropriate dimensionality on the left hand side which we have set to 1, in the BWBm case this is simply the familiar diffusion coefficient.}
\end{table}

In Table 1 we  collect the kinetic equations for the pdf $p=p(x,t)$ of some processes which have been proposed in the various literatures on time series analysis and anomalous diffusion. For clarity we concentrate on the simplest examples from the family of stable processes and from the CTRW. In the table the (statistical) self similarity exponent $H$ is defined using dilation in time where $\Delta t$ goes to $\lambda \Delta t$: 
\begin{equation}
x(\lambda \Delta t) = \lambda^H x(\Delta t)
\end{equation}
and the equality is in distributions.

The fourth row of Table 1 corresponds to Bachelier-Wiener Brownian motion (BWBm) and the fifth row to the familiar diffusion equation where we have abbreviated $\partial / \partial t$  to $\partial_t$. BWBm is of course a manifestation of the central limit theorem (CLT) \cite{Levy1937,Meerschaert2001}. The solution $p(x,t)$ is of Gaussian form with width spreading as $t^{1/2}$ and its characteristic function is also a Gaussian in k $(\sim \exp (-|k|^2 t)$ with stability exponent $\alpha=2$; the process has power spectrum $S(f) \sim f^{-2}$. 

\subsection{Anomalous diffusion and the extended Central Limit Theorem}

Similarly, the fifth row corresponds to relaxing the assumption of finite variance, 
by allowing a stability exponent $0 <  \alpha  < 2$. The corresponding probability density functions (pdf) $p_\alpha(x,t)$ are the $\alpha$-stable class, with power law tails decaying as $x^{-(\alpha+1)}$. Following Mandelbrot we refer to  these as ``L\'{e}vy flights" or ordinary L\'{e}vy motion (oLm). 

The corresponding kinetic equation  
\begin{equation}
\frac{\partial p_{\alpha}(x,t)}{\partial t} = \frac{\partial^{\alpha} p_{\alpha}(x,t)}{\partial |x|^{\alpha}} \label{OLM_KE}
\end{equation}
has a symmetric Reisz fractional derivative in space, $\partial^{\alpha}/\partial |x|^{\alpha}$, which in the Table is given as three dimensional and abbreviated to  $\nabla^{\alpha}$. 
The Reisz derivative is a pseudodifferential operator with symbol $-|k|^{\alpha}$ and $p_{\alpha}(x,t)$ has characteristic function $\hat{p}_\alpha(k,t)= \exp(-|k|^{\alpha}t)$.
Unlike the cases we now go on to discuss, the kinetic equation for oLm is still unambiguously Markovian and  an expression of the extended CLT. Due to infinite divisibility, in this specific case the pdf alone $p_{\alpha}(x,t)$ is enough to uniquely characterise the stochastic process, which we will call $Z_{\alpha}(t)$.

\subsection{Relaxing independence through temporal memory: fractional Brownian motion vs. the fractional time process}

The seventh row of Table 1 describes the case  when the iid assumption is relaxed, rather than the finite variance one. This case is more subtle than the previous two. Relaxing independence is one way to break the iid assumption and is the situation we consider. It can be done in several ways, we will discuss just two.

One of the ways which has been employed in the CTRW formalism is to take a power 
law pdf of waiting times  $p(\tau) \sim \tau^{-(1+\alpha H)}$\citep{LindenbergWest1986}. This became known as the fractional time  process (ftp, see also \cite{Lutz2001}).  The waiting times themselves are still iid, but their infinite mean  is assumed to be a consequence of dependence due to  microscale physics. The kinetic equation  that corresponds to the ftp  \citep{Balakrishnan1986, MetzlerKlafter2000,Lutz2001} can be seen in the fourth column of the eighth row in Table 1. We may define a temporal exponent   by $d^{'}=\alpha H$. 
The fractional derivative in time, of order $\alpha H=d^{'}$, corresponds physically to the power law in waiting times. The prime   indicates that this exponent is not  identical  to the memory parameter $d$ in the case of fBm or FARIMA \cite{BurneckiEA2008}. $d^{'}$ runs from $0$ to $1$ and is, for example, the same as the  temporal exponent defined by    Brockmann et al   (their ``$\alpha$"; our $\alpha$ is their ``$\beta$"). In all the following cases $D$ is no longer the Brownian diffusion constant but simply ensures dimensional correctness in a given equation. Note that we do not include the term describing the power law decay of the initial value here or in subsequent CTRW equations (it is retained and discussed in \cite{MetzlerKlafter2000}, see their eqn. 40).  

Another way to relax independence is to introduce global long range dependence, as pioneered  by  \citet{MandelbrotVanNess1968}. They used a self-affine process with a memory kernel, originally due to Kolmogorov and called by them ``fractional Brownian motion".
Contradictory statements exist in the physics literature concerning the equivalent kinetic equation for fBm corresponding to that for ftp. It has somtimes been asserted  \citep{ZaslavskyBook,WatkinsEA2005} that  the equations are the same, while ftp has sometimes been labelled ``fBm" (c.f. the supplementary material in \cite{BrockmannEA2006}). However the solution of the equation for ftp is now known  to be   non-Gaussian  \cite{MetzlerKlafter2000}, and can be given in terms of Fox functions. Conversely the pdf of fBm is by definition \citep{Mandelbrot1982,McCauley2004} Gaussian but with a variance which ``stretches" with time as $t^{2H}$. The correct kinetic equation for fBm must thus \cite{Lutz2001} be local in time. It is shown in row 8, column 3 of Table 1. Given, to our knowledge, first by \citet{WangLung1990}, it can be seen by trial solution to have a solution of the required form.

The difference between  ftp and fBm is striking, in  that although both include temporal correlations, the kinetic equation for the ftp is non-local in time while that for fBm is local. This distinction disappears when we go to a Langevin description, where both processes explicitly require fractional derivatives \cite{Lutz2001}.  We are grateful to our referee for pointing out that the kinetic equation for fBm also corresponds to a transformation of time to $t^{2 H}$ in the ordinary diffusion equation for BWBm, which we may contrast with  the fractional derivative in time in the kinetic equation for ftp.  We remark that if one rescales BWBm with time, the resulting increments would not be stationary whereas 
fBm with the same kinetic equation  has stationary increments. This illustrates the point that fBm shares its kinetic equation with several other stochastic processes and so a full specification of the process thus requires more than the kinetic equation.

\subsection{Combining memory with infinite variance: ``ambivalence" vs. lfsm.}

\subsubsection{``ambivalent processes" and the fully fractional continuous time random walk}

Questions similar to those in the previous section have been asked in the physics literature about the natural generalisation of the fractional time process to allow for both L\'{e}vy distributions of jump lengths as well as  power-law distributed waiting times. The resulting fully fractional kinetic equation  is  fractional in both  in space and time: 
\begin{equation}
\frac {\partial^{\alpha H}}{\partial t^{\alpha H} } P_{ap}(x,t) = D \frac{\partial^{\alpha}}{\partial |x|^{\alpha}} P_{ap}(x,t)  
\end{equation}
and was used by  \citet{BrockmannEA2006} to exemplify the ``ambivalent process". Analogously  with the ftp  the solution for this process is known \citep{Kolokoltsov} not to be a L\'{e}vy-stable (or stretched L\'{e}vy-stable) distribution but rather a convolution of such distributions. 

\subsubsection{A self-similar stable alternative to the a.p.: Linear fractional stable motion}

Analogously to the generalisation of the ftp to the ambivalent process, there are several $H$-self similar Levy symetric $\alpha$-stable processes, described in \cite{WeronEA2005}. We here consider  the simplest one, linear fractional stable motion (lfsm), which generalises linear fractional Brownian motion to the infinite variance case. We emphasise that lfsm is, for example, not the fractional Levy motion referred to by \citep{Huillet1999}. We can describe lfsm through a stochastic integral:

\begin{equation}
S_{H}(t|\alpha,b_1,b_2)=\int_{-\infty}^{\infty} K_{H,\alpha}(t-s) Z_{\alpha}(ds)
\end{equation}  
where the memory kernel $K_{H,\alpha}$ is given by
\begin{eqnarray}
K_{H,\alpha} & = & b_1 \large[ (t-s)^{H-1/\alpha}_{+} - (-s)^{H-1/\alpha}_{+} \large]\nonumber \\
             & + & b_2 \large[ (t-s)^{H-1/\alpha}_{-} - (-s)^{H-1/\alpha}_{-} \large]
\end{eqnarray}

\citet{BurneckiEA1997} showed how mixed linear fractional stable motion can be obtained from $Z_{\alpha}(t)$ using the Lamperti transformation \cite{Lamperti1962,EmbrechtsMaejima2002}, a more general result which enables any self-similar process to be obtained from its corresponding stationary stochastic process. We are concerned here, however, simply with obtaining the kinetic equation.  This can be found by direct differentiation of the characteristic function with respect to time (c.f. \cite{PaulBaschnagel1999}).

As with the simpler stable processes the pdf $p_{lfsm}$ of lfsm  can  be expressed via the Fourier transform of the characteristic function (e.g. \cite{SamTaqBook,LaskinEA2002})
\begin{equation}
p_{lfsm}=\int e^{ikx} \exp (-\bar{\sigma} |k|^{\alpha} t^{\alpha H})  \label{LFSM}
\end{equation} 
We  see that the characteristic function: $\hat{p}= \exp (-\bar{\sigma} |k|^{\alpha} t^{\alpha H})$
generalises the oLm case.  Because $\alpha$ is no longer equal to $1/H$  the effective width parameter now grows like $t^{\alpha H}$. The characteristic function has the correct fBm limit, when  $\alpha=$  2, we see for fBm  at any given $t$ it is a Gaussian with width growing as $t^{2H}$. We can see that lfsm is a general stable self-affine process  by taking $k'=k\tau^H$ which gives
$p_{lfsm}=t^{-H} \phi_{\alpha}(x/t^H)$,  a stable distribution of index $\alpha$ and a prefactor ensuring $H-$selfsimilarity in time \cite{KrishnamurthyEA2000}.

Direct differentiation of this pdf gives
\begin{equation}
\frac{\partial}{\partial t} p_{lfsm} = - \alpha H \bar{\sigma} t^{\alpha H -1} \int_{-\infty}^{\infty} e^{ikx} |k|^{\alpha} \exp (-\bar{\sigma} |k|^{\alpha} t^{\alpha H})
\end{equation}
which, absorbing the constant $\bar{\sigma}$, and factors of $2\pi$ into $D$ can be recognised as
\begin{equation}
\frac{\partial }{\partial t} p_{lfsm} = \alpha H t^{\alpha H-1} D
\frac{\partial^{\alpha} }{\partial x^{\alpha}} p_{lfsm} \label{LFSM_KE}
\end{equation}
using the above definition of the Reisz derivative. Surprisingly the kinetic equation of lfsm seems not have been given explicitly before in either the physics or mathematics literature. \citet{KrishnamurthyEA2000}  quoted an equation of motion for integrated activity in lfsm. This  has a more complicated structure presumably due to additional memory effects arising from the integration process.
 
We note that $\alpha H t^{\alpha H-1}=\partial_t t^{\alpha H}$. This factor arises because  (\ref{LFSM_KE}) could also be obtained from the space  fractional diffusion equation (\ref{OLM_KE}) by a simple transformation of the time variable: $t$ is replaced by $t^{\alpha H}$. The appropriate limits may be easily checked; in particular $\alpha=2$ gives the kinetic equation of fBm. 

We also remark that lfsm should be a special case of the nonlinear shot noise process studied by \citet{EliazarKlafter2006} which may allow further generalisation of the kinetic equation we have presented.    

\section{lfsm as a model of intermittent bursts}

Intermittency is a frequently observed property in complex systems, and can be studied within several paradigms. One such, of continuing interest, has been Bak et al's self-organized criticality (SOC), a key postulate of which is that slowly driven, interaction dominated, thresholded dynamical systems will establish  long ranged correlations via ``avalanches" of spatiotemporal activity. The avalanches are found to obey power laws in size and duration. In consequence, many papers have sought to measure ``bursts" of activity in natural time series. This has most typically been done by means of a fixed threshold. The  duration $\tau$ and size $s$ of the bursts are then respectively defined as the interval  between the i-th upward crossing time ($t_i$) and the next downward crossing time ($t_{i+1}$) of the threshold, and the integrated area above the threshold between these times.  

The search for SOC in the magnetosphere and solar wind has used this approach among others (e.g. \cite{FreemanEA2000,Watkins2002}).  The similarity of observed burst size and duration distributions in solar wind and magnetospheric quantitities to those from models of turbulence and SOC led \citet{FreemanEA2000} and \citet{Watkins2002} to conjecture that, at least qualitatively, such behaviour might   simply be an artifact of a self similar (or multifractal) time series, rather than unique to a given mechanism. In particular this was in distinction to the idea that one could use the presence or absence of power laws in waiting times defined similarly to the above as a distinguishing feature between SOC and turbulence. One of the present authors has thus elsewhere \cite{Watkins2002} advocated the testing of avalanche diagnostics using controllable self similar models. Similar points have been made subsequently by \citet{CarboneEA2004} for fBm and \citet{Bartolozzi2007} for the multifractal random walk.   

In this section we thus present a preliminary investigation of the ability of lfsm to qualitatively mimic SOC signatures in data. As the kinetic equation we have derived is not unique to lfsm and   is insufficient to specify all its  properties in what follows  we have used a numerical simulation of the process $S_{}$ using  the algorithm of \citet{StoevTaqqu2004} and analytic arguments, after those of  \citet{KearneyMajumdar2005} to predict the scaling of the tail of the pdf of burst size $s$ and duration $\tau$ in lfsm for large $s,\tau$. Rather than estimate the exponents from plots of the pdf or empirical complementary cdf we have elected to use maximum likelihood estimation, as implemented in the algorithms of A. Clauset and co-workers (http://www.santafe.edu/~aaronc/powerlaws/). Detailed comparison with measured experimental exponents is not attempted at this stage and will be the subject of future work. 
 
\begin{figure}
\label{figure1}
\begin{center}

\includegraphics*[width=9cm,angle=0]{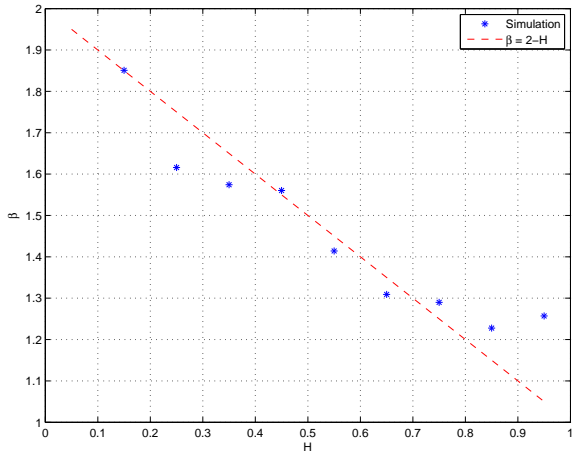}

\end{center}
\caption{Dependance of exponent $\beta$ for pdf  $p(\tau)$ of a burst of duration $\tau$ on $H$ for simulated lfsm in the fBm, $\alpha=2$ limit. The average of 7 trials was taken.}
\end{figure}

The expected behaviour of burst duration and sizes for lfsm has, as far as we know, not been investigated. 
Dealing first with durations,  we make use of the fact that for a fractal curve of self similarity exponent $H$ and dimension $D=2-H$ the points $\{ t_i \}$ have dimension $1-H$. In consequence the probability of crossings over a time interval $\tau$ goes as $\tau^{1-H}$ giving an inter-event probability scaling like $\tau^{-(1-H)}$. The pdf for inter-event intervals in the isoset thus scales as: 
\begin{equation}
p(\tau) \sim \tau^{\beta} \label{pdf_beta}
\end{equation}
 where
\begin{equation}
\beta=2-H \label{exponent_beta}
\end{equation} 
  giving the same exponent $3/2$ as for the first passage distribution in the Brownian case. 
For symmetric processes this scaling is retained by the subset of the isoset that corresponds to burst ``durations". We expect this to be independent of the detailed nature of the model and so should, in particular, also apply to lfsm.  

 To establish the behaviour of burst ``sizes" we note first that \citet{KearneyMajumdar2005}  considered the zero-drift Wiener Brownian motion (BWBm) case. Rather than their full analytic treatment, we recap their heuristic argument for a burst size (area) $A$ defined using the first-passage time $t_f$. This may then be adapted to burst sizes defined using isosets, and thence to lfsm. They first noted that for BWBm the instantaneous value of the random walk $y(t)\sim t^{1/2}$ for large $t$. Then, defining $A$ by
\begin{equation}
A=\int_{t_i}^{t_f} y(t') dt'
\end{equation}
the integration implies that large $A$ scales as $t_f^{3/2}$. Simple inversion of this expression implies that $t_f$ must scale as $ t_f \sim A^{2/3}$. We independently have the standard result for first passage time for BWBm:
$P(t_f) \sim t_f^{-3/2}$. To get $P(t_f)$ as a function of $A$ i.e. $P(t_f(A))$  one needs to insert the expression for $t_f$ as a function of $A$ in above equation, and in addition will need a Jacobian.  After these manipulations  $P(A)\sim A^{-4/3}$  \cite{KearneyMajumdar2005}.
 
In the zero-drift but non-Brownian case we will still argue that $y(t)\sim t^{H}$ for large $t$. As our application uses the above mentioned isoset-based burst size $s$ rather than those based on first passage times we define
$s$ by: 
\begin{equation}
s=\int_{t_i}^{t_{i+1}} y(t') dt'
\end{equation}
 
The rest of the argument goes as before but using (\ref{pdf_beta}).  We find:
\begin{equation}
P(s) \sim s^{\gamma}\label{pdf_gamma}
\end{equation}
where 
\begin{equation}
\gamma=-2/(1+H)\label{exponent_gamma}
\end{equation}
which we can  check in   the Brownian case where $H=1/2$ to retrieve $P(s) \sim s^{-4/3}$.   

The same exponents, $\beta$ and $\gamma$, but defining the bursts using a DFA-like moving average rather than a fixed threshold, were earlier  investigated, for the fBm case only, by \citet{CarboneEA2004}. The format of our figures for the fBm and lfsm cases has been chosen to allow  comparison with theirs. They found numerically the same dependences of $\beta$ and $\gamma$ on H, as we have in equations (\ref{exponent_beta}) and (\ref{exponent_gamma}) above, which is intuitively reasonable with hindsight because the choice of fixed or running threshold should not change the asymptotic scaling behaviour.  For a fixed threshold the burst size and duration exponents for fBm have also very recently been presented by \cite{Rypdal2008}.  

Our numerics confirm that using the fixed threshold definition the expressions for $\beta$ and $\gamma$ also describe fBm well, although the scatter, from a single trial in the case of each value of $\beta$ and $\gamma$ shown in Figures 3 and 4, seems relatively high. We have reduced the scatter in figures 1 and 2 by plotting the average of the exponents over a small number of trials (here 7). The assumption that burst size $s$ grows with duration $\tau$ used in our heuristic derivation above can be seen to be reasonable for the fixed threshold, fBm case in Figure 5.

Perhaps more surprisingly the expressions also seem to hold  reasonably well when the stability exponent is reduced, first to 1.8 (Figures 6 and 7) and then to 1.6 (Figures 8 and 9). Again we note that these are averages of 7 trials in each case. By the $\alpha=1$ case presented in Figures 10 and 11, however, the expressions can be seen to fail. In this parameter regime, for any given $H$, they are seen to consistently underestimate both the burst exponents. It has been suggested to us that this could be because $y^H$ ceases to be a good estimate of characteristic displacement when the increments of the walk are very heavy tailed (S. Majumdar, Personal Communication, 2006) but we have so far been unable to find a suitable alternative expression. 

\begin{figure}
\label{figure2}
\begin{center}
\includegraphics*[width=9cm,angle=0]{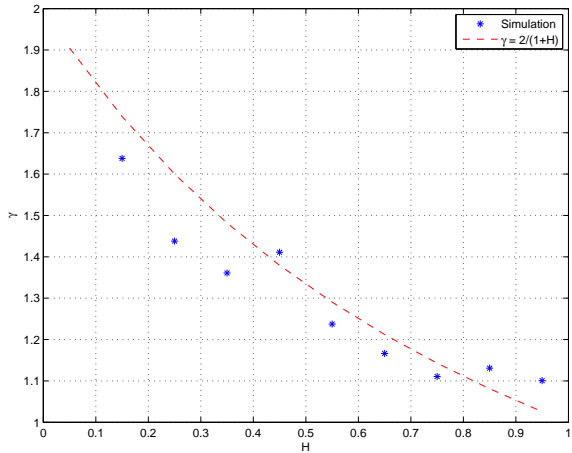}
\end{center}
\caption{Dependance of exponent $\gamma$ for the pdf  $p(s)$ of a burst of size $s$ on $H$ for simulated lfsm in fBm limit. Again an average of 7 trials was taken.}
\end{figure}

\begin{figure}
\label{figure3}
\begin{center}
\includegraphics*[width=9cm,angle=0]{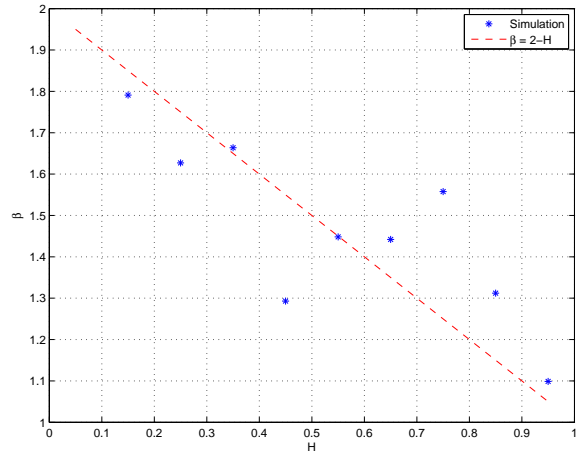}
\end{center}
\caption{As Figure 1, dependence of duration exponent $\beta$ on H, but 1 trial only.}
\end{figure}

\begin{figure}
\label{figure4}
\begin{center}
\includegraphics*[width=9cm,angle=0]{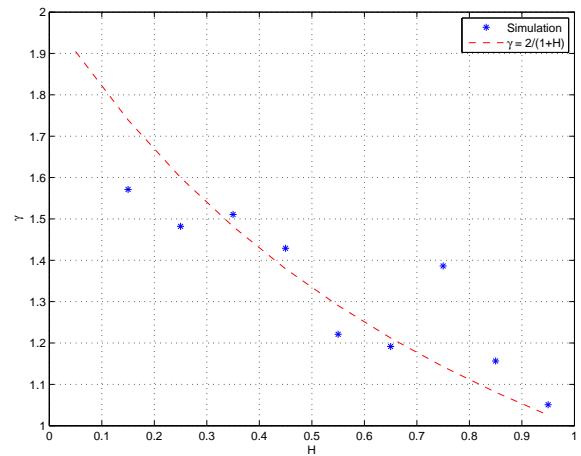}
\end{center}
\caption{As Figure 2, dependence of size exponent $\gamma$ on H, again  1 trial only.}
\end{figure}

\begin{figure}
\label{figure5}
\begin{center}
\includegraphics*[width=9cm,angle=0]{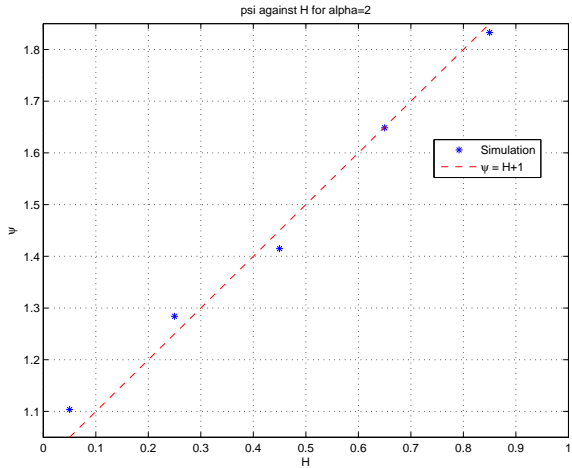}
\end{center}
\caption{Dependence of  exponent $\psi$ on $H $in fBm case. $\psi$ captures growth of burst size $s$ with duration $\tau$}
\end{figure}

\begin{figure}
\label{figure2}
\begin{center}
\includegraphics*[width=9cm,angle=0]{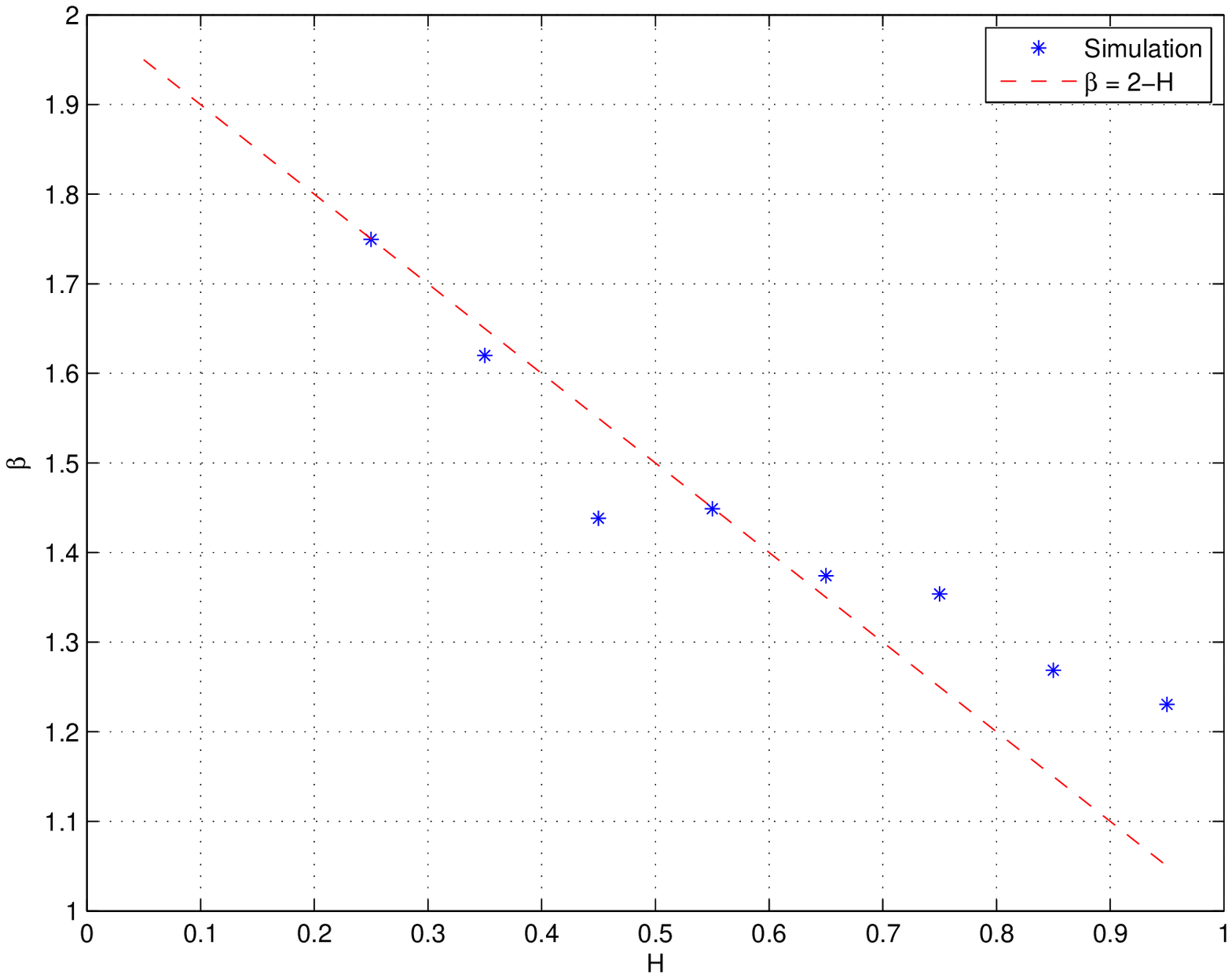}
\end{center}
\caption{As Figure 1 (burst duration exponent $\beta$ vs H, 7 trials), but for $\alpha=1.8$. }
\end{figure}

\begin{figure}
\label{figure2}
\begin{center}
\includegraphics*[width=9cm,angle=0]{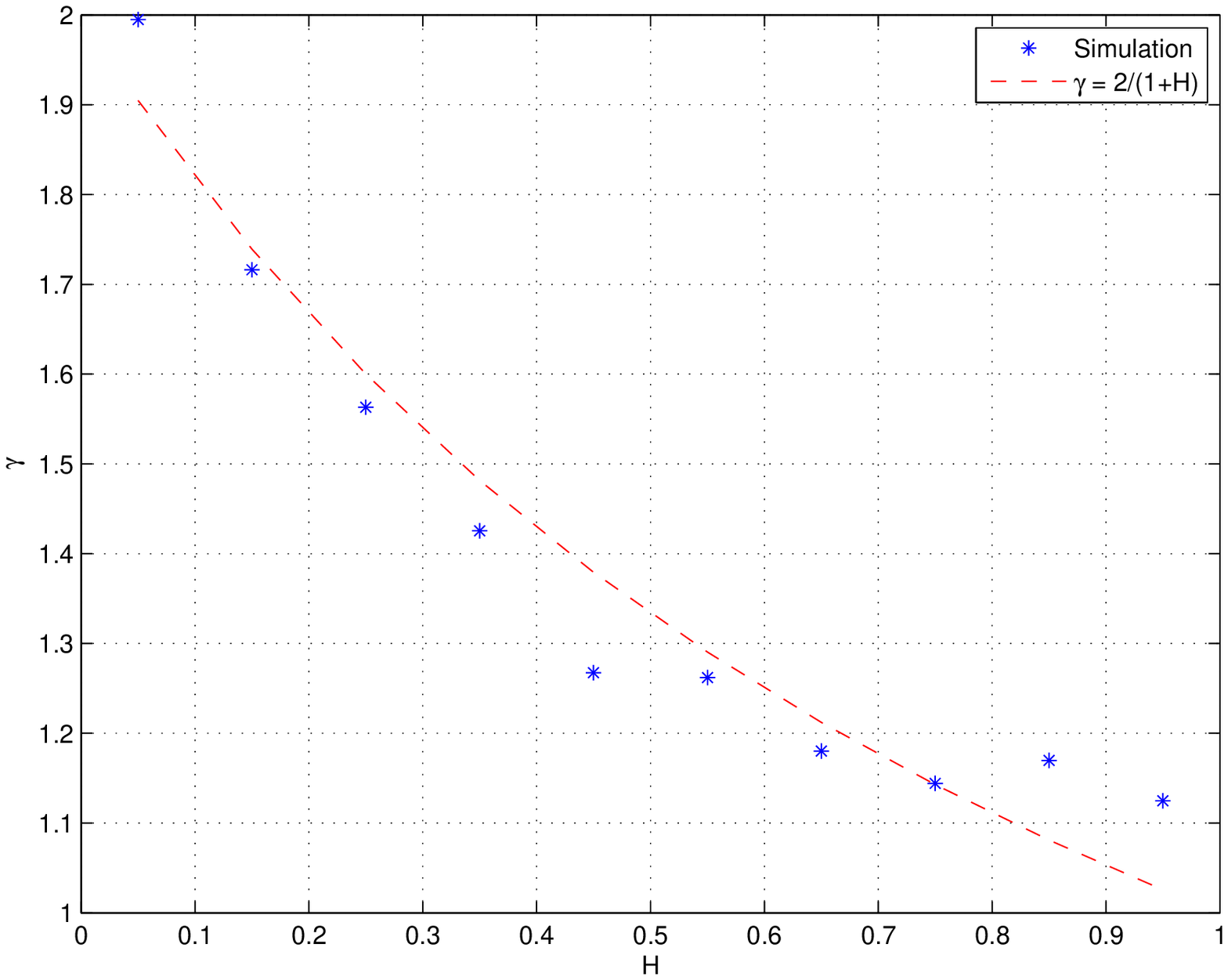}
\end{center}
\caption{As Figure 2 (burst size exponent $\gamma$ vs H, 7 trials), but for $\alpha=1.8$. }
\end{figure}

\begin{figure}
\label{figure2}
\begin{center}
\includegraphics*[width=9cm,angle=0]{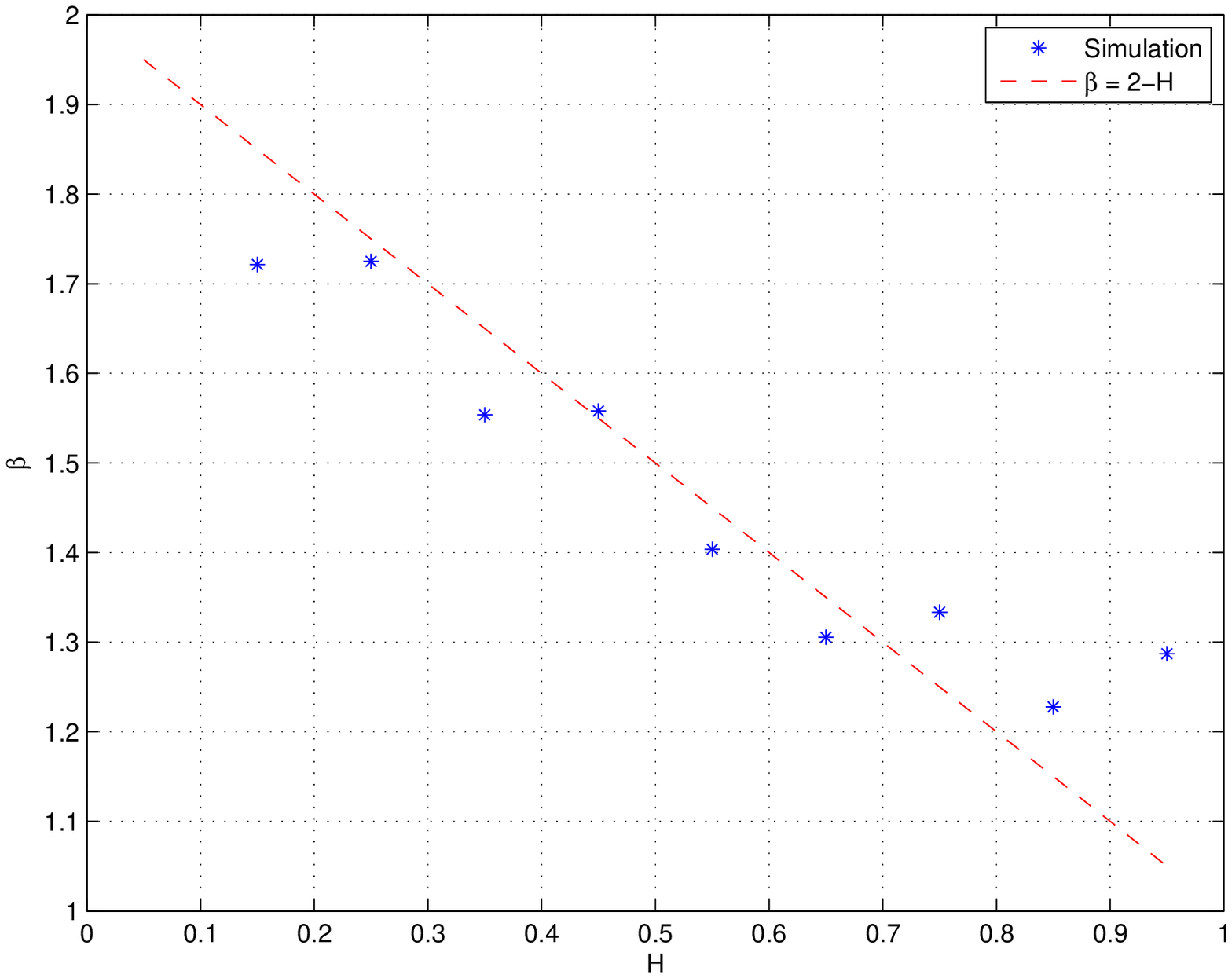}
\end{center}
\caption{As Figure 1 (burst duration exponent $\beta$ vs H, 7 trials), but for $\alpha=1.6$.}
\end{figure}

\begin{figure}
\label{figure2}
\begin{center}
\includegraphics*[width=9cm,angle=0]{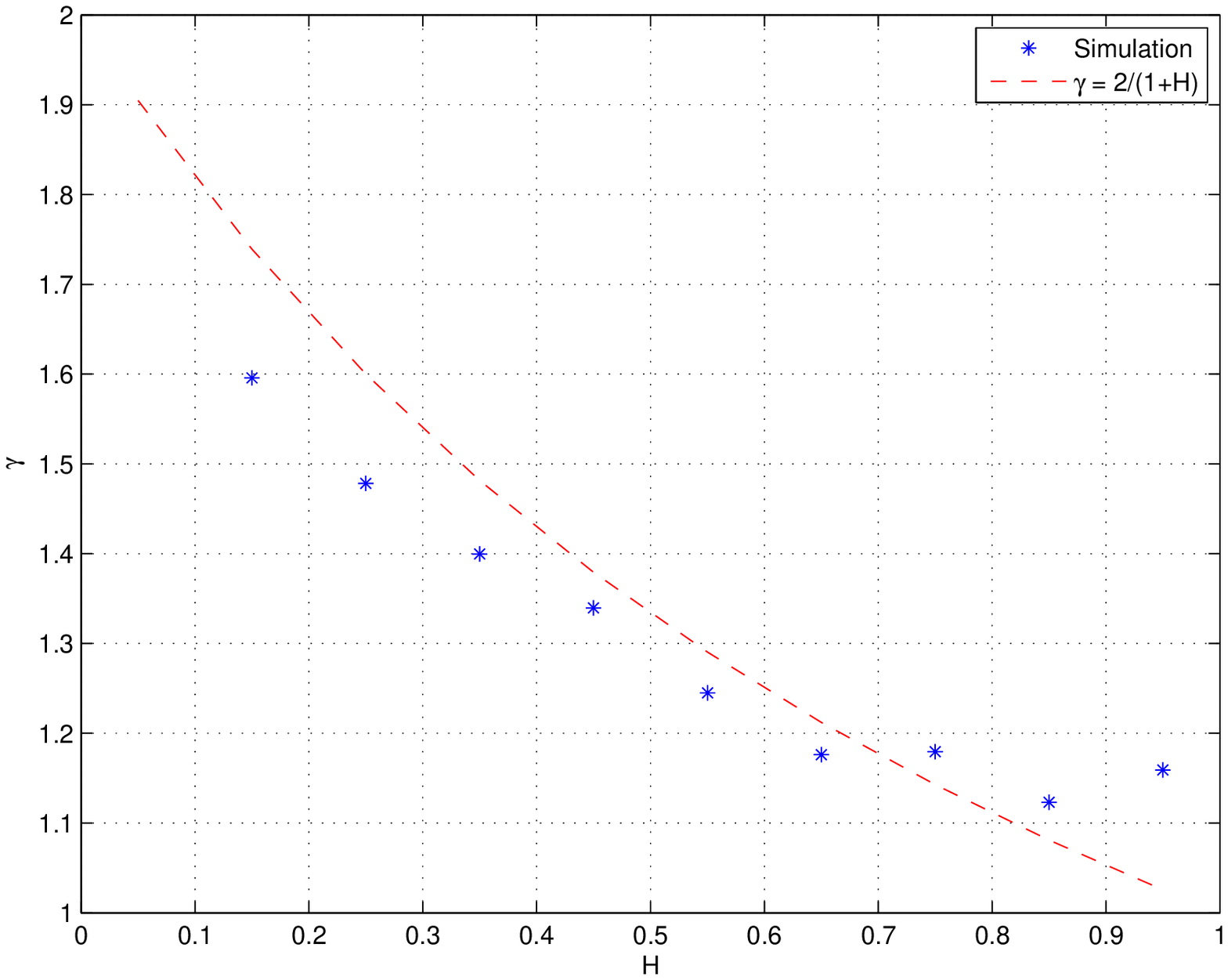}
\end{center}
\caption{As  Figure 2 (burst size exponent $\gamma$ vs H, 7 trials), but for $\alpha=1.6$ }
\end{figure}

\begin{figure}
\label{figure2}
\begin{center}
\includegraphics*[width=9cm,angle=0]{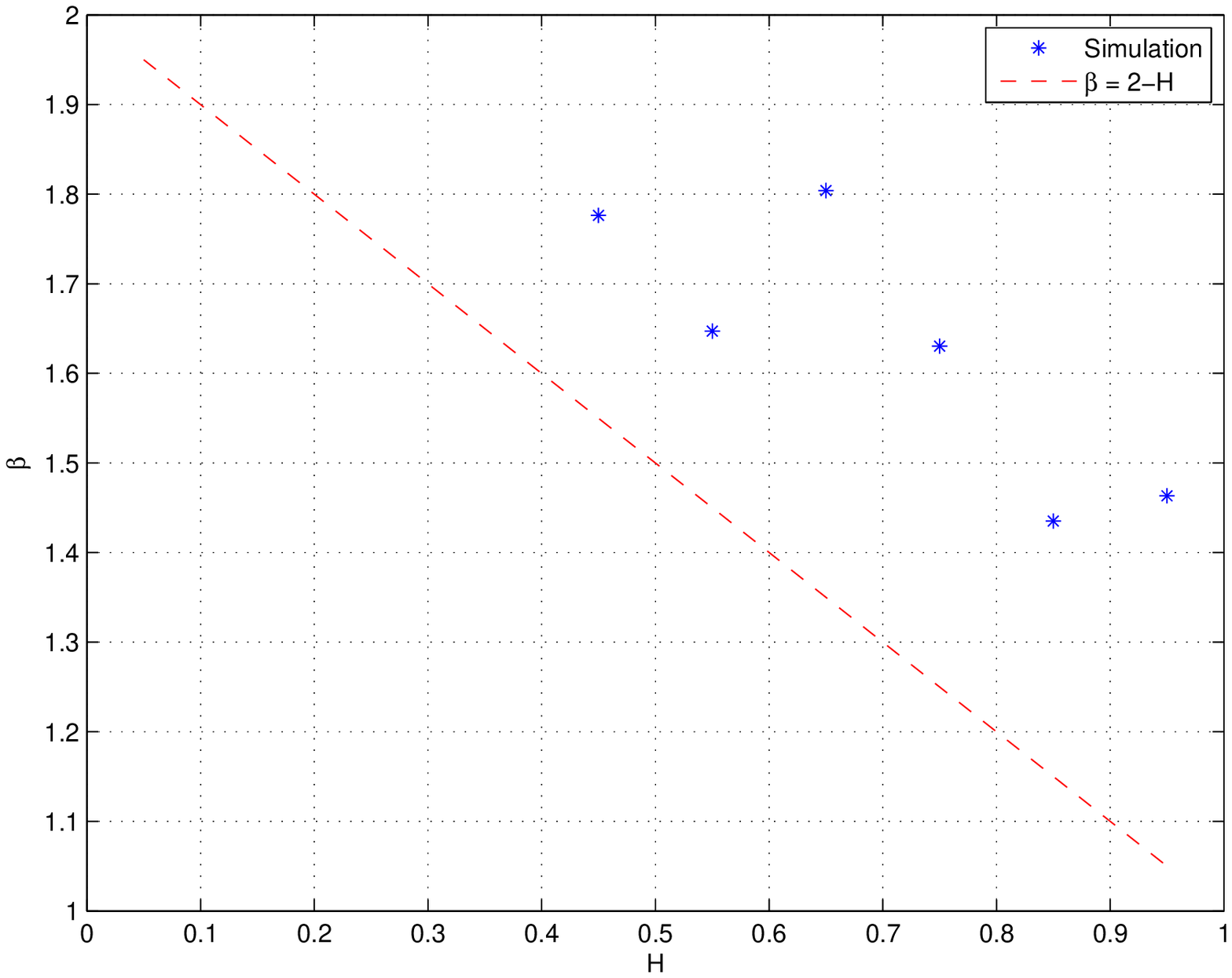}
\end{center}
\caption{As  Figure 1 (burst duration exponent $\beta$ vs H, 7 trials), but for $\alpha=1$.}
\end{figure}

\begin{figure}
\label{figure2}
\begin{center}
\includegraphics*[width=9cm,angle=0]{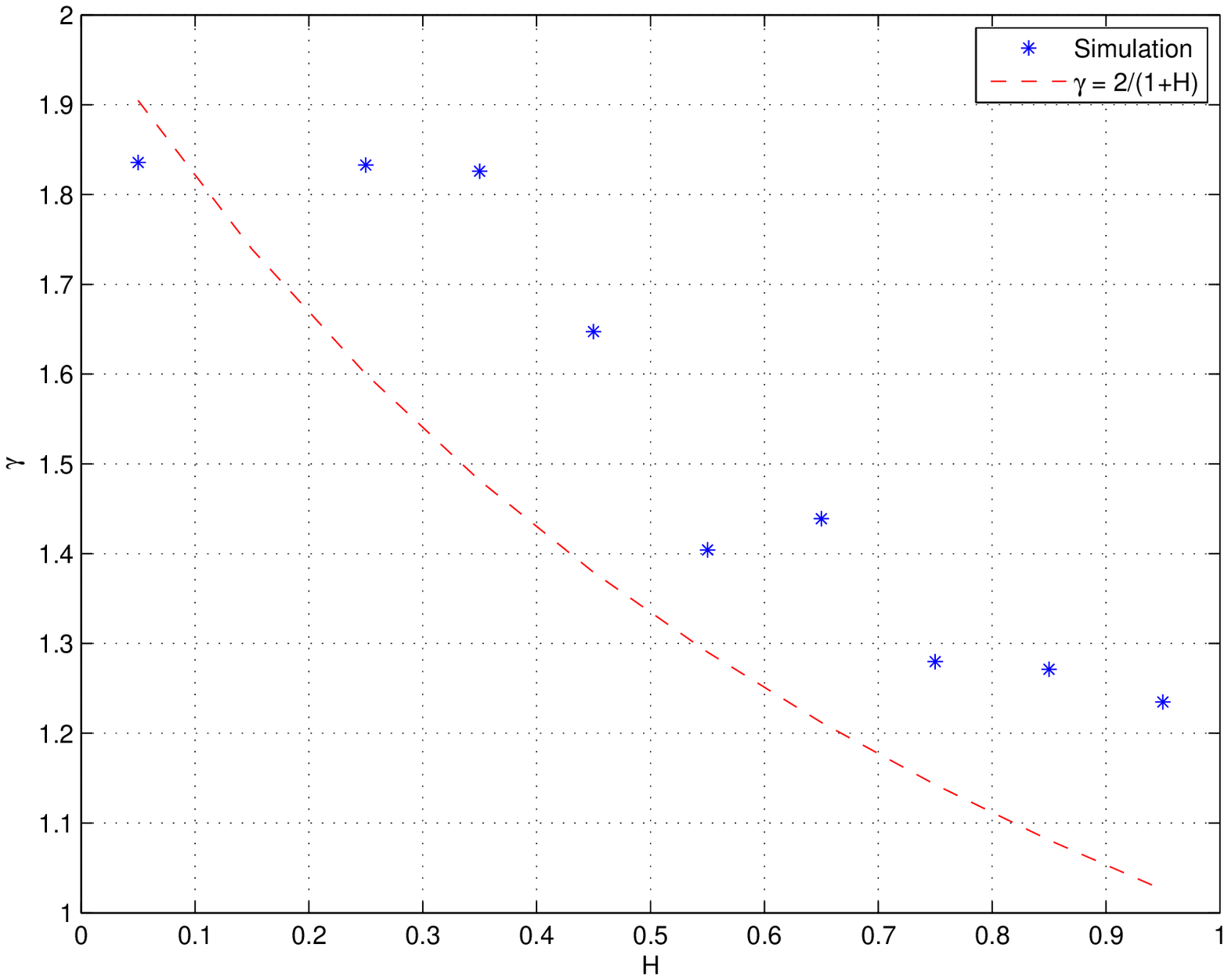}
\end{center}
\caption{As Figure 2 (burst size exponent $\gamma$ vs H, 7 trials), but for  $\alpha=1$ }
\end{figure}

\section{Conclusions} 

In this paper we studied the question of whether one would expect the same equation to describe a time series as an anomalous diffusive process. A codification of diffusion-like equations showed that a kinetic equation was ``missing" from the literature; the one corresponding to lfsm. We gave a simple derivation for it by direct differentiation of the well-known characteristic function of lfsm.  We then made a preliminary exploration of how lfsm could model the   ``burst" sizes and durations  previously measured on magnetospheric  and solar wind time series.  We made simple scaling arguments building on a result of \citet{KearneyMajumdar2005} to show how lfsm could  be one candidate explanation for  such ``apparent SOC" behaviour and made preliminary comparison  with numerics. These arguments fail when the tails of the pdf of increments become very heavy, and further work is needed on this topic.  

In future we also plan to consider other stochastic processes, both  FARIMA (c.f. \citep{BurneckiEA2008}) and nonlinear shot noises, to allow generalisation of the above initial investigations into burst size and duration. The prevalence of natural processes showing heavy tails and/or long ranged persistence suggests a relevance well beyond our initial area of application in space physics.

We thank in particular Alex Weron, Kristof Burnecki and Marcin Magdziarz for many helpful comments on earlier versions of the paper. NWW is also grateful to Mikko Alava, Robin Ball, Tom Chang, Aleksei Chechkin, Joern Davidsen, Mervyn Freeman, Bogdan Hnat, Mike Kearney, Khurom Kiyani, Yossi Klafter, Vassili Kolokoltsev, Eric Lutz, Satya Majumdar, Martin Rypdal, Dimitri Vyushin and Lev Zelenyi for valuable interactions. NWW acknowledges the stimulating environments of the Newton Institute programme PDS03  and the KITP programme  ``The Physics of Climate Change". Research was carried out in part at Oak Ridge National Laboratory, managed by UT-Battelle, LLC, for U.S. DOE under contract number DE-AC05-00OR22725. This research was supported in part by the EPSRC, STFC and NSF under grant number NSF PHY05-51164.

\end{document}